\documentclass[prl,twocolumn,showpacs]{revtex4}

\usepackage[dvips]{graphics}
\usepackage{amsmath, amsthm, amssymb}
\usepackage{graphicx}
\usepackage{dcolumn}
\usepackage{bm}

\newcommand{\de}[1]{\left( #1 \right)}

\newcommand{\ket}[1]{\left| #1 \right\rangle}
\newcommand{\bra}[1]{\left\langle #1 \right|}
\newcommand{\braket}[2]{\left\langle #1 \mid #2 \right\rangle}

\newcommand{\mean}[1]{\left\langle #1 \right\rangle}

\newcommand{\tr}{\mathrm{Tr}}

\newcommand{\ie}{{\it{i.e.}}}
\newcommand{\etal}{{\it{et al.}}}

\begin{document}

\title{Geometric Phase Transition: direct measurement of a mathematical abstraction}

\author{Daniel Cavalcanti}\email{dcs@fisica.ufmg.br}
\affiliation{Departamento de F\'{\i}sica - Caixa Postal 702 -
Universidade Federal de Minas Gerais - 30123-970 - Belo Horizonte -
MG - Brazil}
\author{Fernando G.S.L. Brand\~ao}\email{fernando.brandao@imperial.ac.uk}
\affiliation{QOLS, Blackett Laboratory, Imperial College London,
London SW7 2BW, UK}\affiliation{Institute for Mathematical Sciences,
Imperial College London, London SW7 2BW, UK}
\author{Marcelo O. \surname{Terra Cunha}}\email{tcunha@mat.ufmg.br}
\affiliation{Departamento de Matem\'atica - Caixa Postal 702 -
Universidade Federal de Minas Gerais - 30123-970 - Belo Horizonte -
MG - Brazil}

\begin{abstract}
By viewing entanglement as a state function, a new kind of phase
transition takes place: the {\emph{geometric phase transition}}.
This phenomenon occurs due to singularities in the shape of the
entangled states set. It is shown how this result can be carried to
provide a better understanding of the geometry of entanglement.
Surprisingly this study can be done experimentally, what allows to
determine the shape of different entangled states sets, a purely
mathematical definition, in real experiments.
\end{abstract}


\maketitle

Phase transition is a general phenomenon which can occur in
different physical systems. It is characterized by a singular
behavior of a state function under certain conditions \cite{LL}. The
standard example is the liquid-gas transition in fluids where a
discontinuity in the density happens when the system reaches a
critical temperature.
 Analogous situations can happen in
quantum systems, a phenomenon called Quantum Phase Transitions
\cite{Sac}. It has been widely discussed that
entanglement - exclusively quantum correlations \cite{Bell} - 
plays a major role in quantum phase transitions
 \cite{ON,WSL,V,O,Nos,VDC} and, in fact, this correlation can be
viewed as a true order parameter in their characterization
\cite{Bra05}.

A great effort has been devoted to understand the relations between
entanglement and critical phenomena in several systems
\cite{ON,WSL,V,O,Nos,VDC,Bra05}. In fact, it is natural to associate
these concepts once correlations are behind both of them
 \cite{Yang}. By sharing this point of view, \ie that entanglement
can be treated as a singular state function and thus as an order
parameter, we obtain interesting novelties such as the appearance of
a new kind of phase transition, that we term ``geometric phase
transition'', concerning a singular behavior of entanglement solely
due to geometric consequences of quantum correlations. It happens
whenever the set of quantum states presents a singular shape. We
also show how this phenomenon allows a more accurate study regarding
the geometry of the set of entangled states. Moreover it is also
shown how this study can be made experimentally by implementing
entanglement-witness operators.

But what means a state to be entangled? This question, although very
clear in the bipartite context, becomes vague in the multipartite
case. When several parts are involved, we have to specify the kind
of entanglement we are interested in. Let us be more specific. For
two systems $A$ and $B$, a state is said to be separable if it can
be written as $\rho=\sum_i p_i \rho^{A}_{i} \otimes \rho^{B}_{i}$
($\{p_i\}$ is a probability distribution), if not, it is said
entangled. However, if a system has $n$ parts, $A$, $B$, $C$,
$\ldots$, $N$, one can talk about entanglement with respect to any
specific partition; moreover, one can also talk about entanglement
with respect to a certain number of parts. For example, take $n=3$.
If one consider $AB$ as a single object, one can define separability
with respect to the partition $AB|C$ exactly as is done for
bipartite systems. Of course, the same can be done with respect to
other bipartitions (only $3$ bipartitions are possible for $n=3$).
Explicitly, a tripartite state is ($AB|C$)-separable if it can be
written as $\displaystyle{\rho _{ABC} = \sum _i p
_i\rho^{AB}_{i}\otimes \rho _{i}^{C}}$. In opposition, any
($AB|C$)-non-separable state has ($AB|C$)-entanglement. If a state
can be written as a convex combination of ($AB|C$)-separable,
($BC|A$)-separable and ($CA|B$)-separable states, then it is said
$2$-separable. By this example, it should become clear the existence
of a multitude of kinds of entanglements for (large $n$)
multipartite systems.

\begin{figure}[ht]
    \includegraphics[scale=0.30]{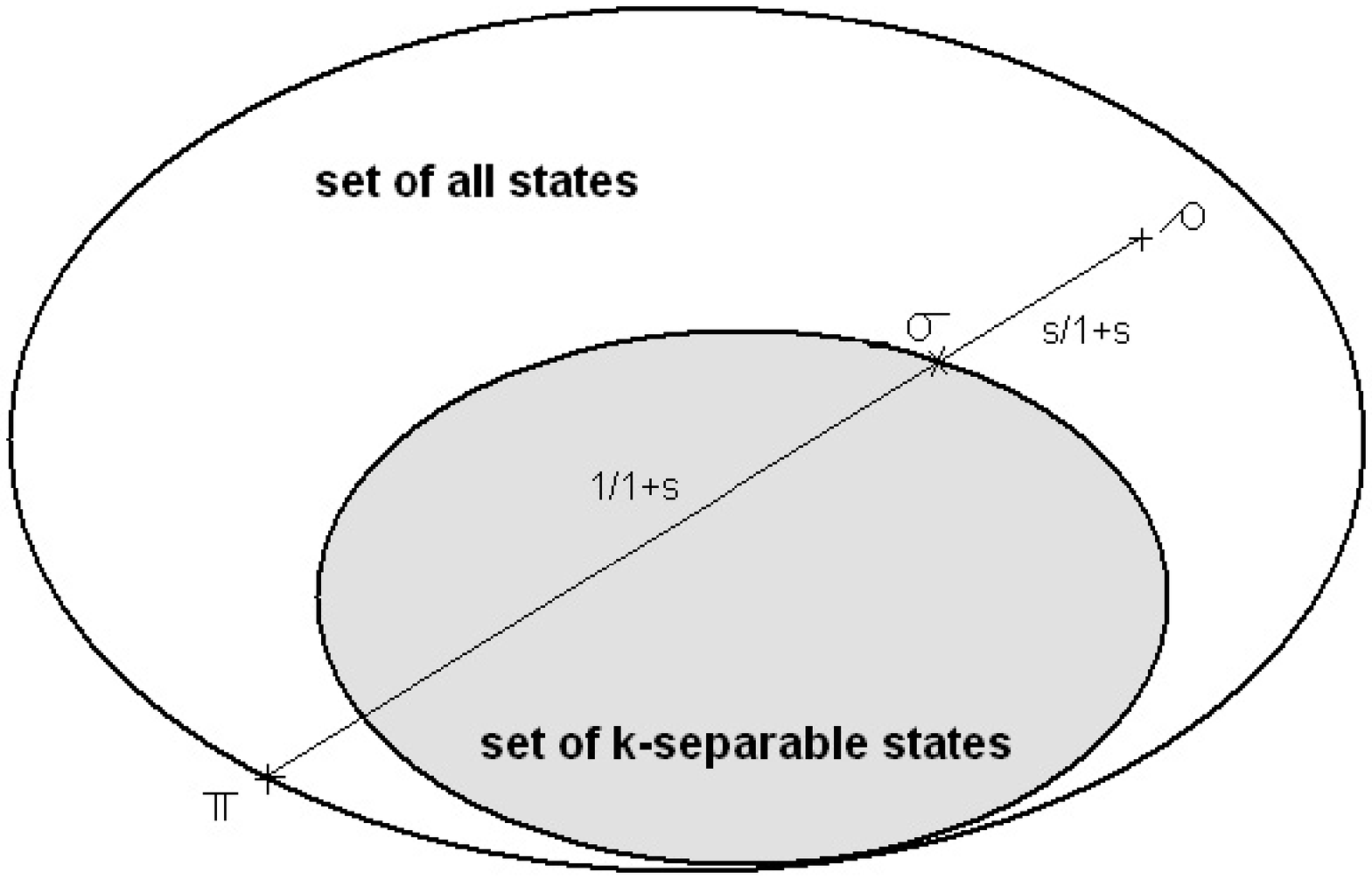}\\
    \includegraphics[scale=0.30]{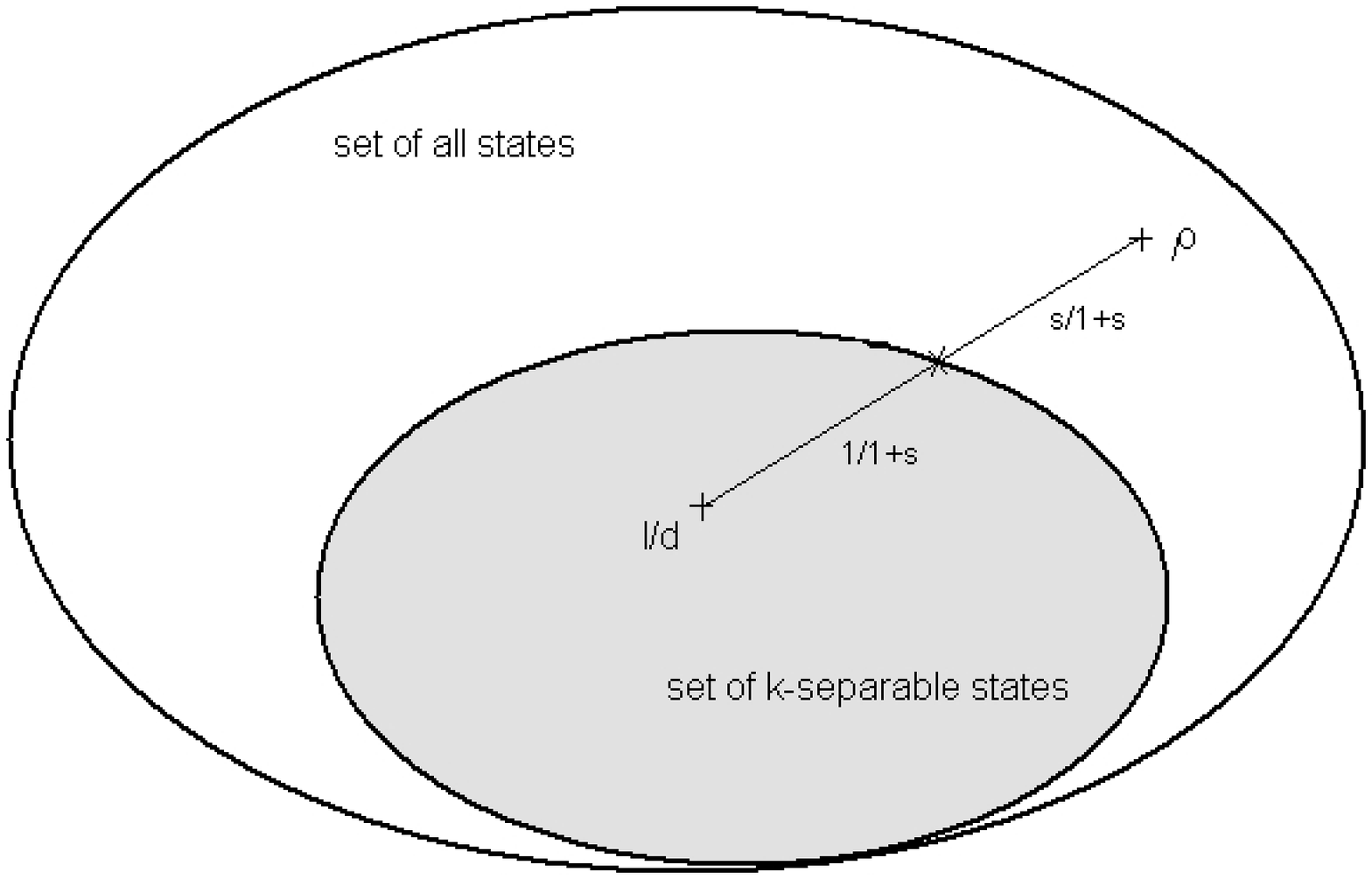}\\
  \caption{The states $\rho$, $\sigma$, and $\pi$ can be represented as points in the space of quantum
  states. A subset of all density matrices is the set of states that
  do not contain a certain kind of entanglement, shown in the
  figure as the $k$-separable states (explained in the text). {\bf{Above:}} The line connecting $\rho$ and $\pi$
  represents the convex combination $\frac{\rho+s\pi}{1+s}$. It is possible to see that, for some choices of $\pi$,
  this
  combination reaches the set of $k$-separable states, for a determined value
  of $s$, in the point $\sigma$. The generalized $k$-robustness, $R_g^k\de{\rho}$ is the minimum
value of $s$ when all possible $\pi$ states
  are considered.
  As the figure indicates, it can be concluded that $\pi$ must be in the
  boundary of the set of all states while $\sigma$ is on the boundary of $S_k$.
   {\bf{Below:}} The difference between $R_g^{k}$ and $R_r^{k}$
  is that in $R_r^{k}$ the state $\pi$ is kept fixed as $I/d$ independently
  of the state $\rho$ to be quantified. As $I/d$ is a separable state, it is a point inside the
  $k$-separable set.}\label{FigRob}
\end{figure}

An important progress in the quantum information field was to note
that entanglement can be treated as a useful resource in various
tasks such as cryptography, quantum computation, and teleportation
\cite{NC}. Thus, like any physical resource, it would be interesting
to properly quantify it. However this goal was not achieved yet, in
spite of the large number of interesting quantifiers already
proposed \cite{expquantif,expquantif2}. A nice example is the
{\emph{robustness of entanglement}} \cite{VT}, which relies on an
interesting geometric interpretation. We should start by defining
the robustness of a state $\rho$ with respect to another state $\pi$
as the minimum $s$ such that the state
\begin{equation}
\sigma=\frac{\rho+s\pi}{1+s}
\label{rob}
\end{equation}
is separable. We will be interested in two special situations. The
first of them, called {\emph{random robustness}} of $\rho$, and
denoted $R_r\de{\rho}$, is obtained when $\pi$ is fixed to be the
completely random state $\frac{I}{d}$, where $I$ is the identity
$d\times d$ matrix. The random state is an interior point in the set
of separable states\cite{Zyc}, which ensures a finite value to
$R_r\de{\rho}$. 
In the second case, we consider the {\emph{generalized robustness}},
denoted $R_g\de{\rho}$, which is obtained by the minimization of the
relative robustness over all states $\pi$ \cite{Steiner}. It follows
$R_g\de{\rho} \leq R_r\de{\rho}$. As we shall see, robustness has
another advantage as an entanglement quantifier: it is generalizable
to multipartite entanglement. The geometrical aspects of those
robustnesses are explained in
 Figure \ref{FigRob}.

Whenever one chooses a kind of entanglement, the set of states which
do not have such entanglement is convex (as a consequence of the
definition of separability). Just to fix ideas, let us talk about
$k$-separability, \ie the states which can be written as convex
combination of states which are product of $k$ tensor factors. The
reader should note that any state is $1$-separable, and that, for a
system of $n$ parts, $n$-separability is separability itself. Let us
denote by $S_k$ the set of $k$-separable states, and by $D=S_1$ the
set of all states. Let us also define the random $k$-robustness of
the state $\rho$, $R^k_r\de{\rho}$, in the same way as before, as
the minimum $s$ such that the state of Eq.~\eqref{rob}, with $\pi =
\frac{I}{d}$, is $k$-separable, and also the generalized
$k$-robustness, $R^k_g\de{\rho}$, as the minimum over all $\pi$
state of the relative $k$-robustness.

\begin{figure}[ht]\centering
    \includegraphics[scale=0.30]{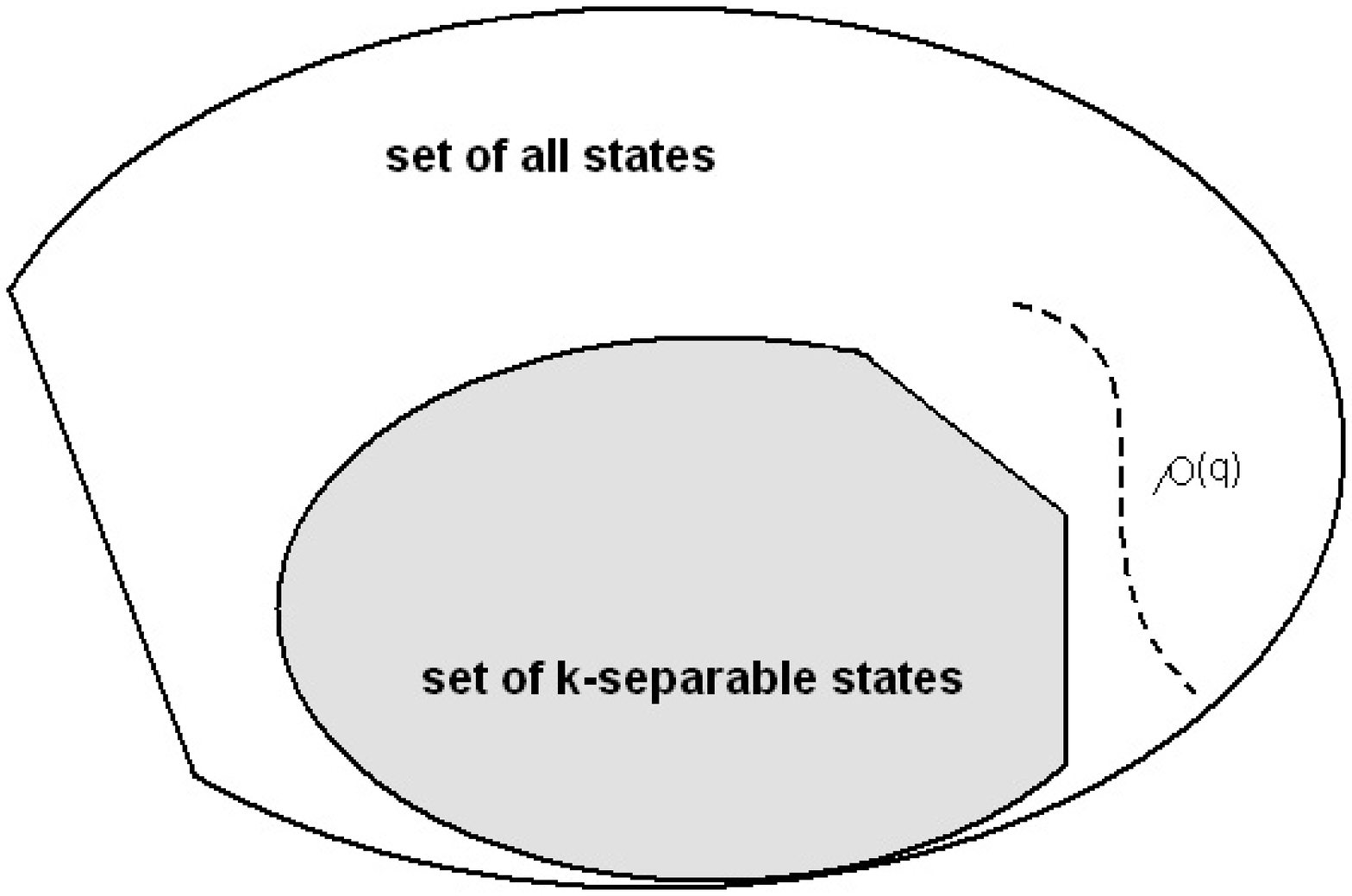}\\
    \begin{tabular}{cc}
    \includegraphics[scale=0.20]{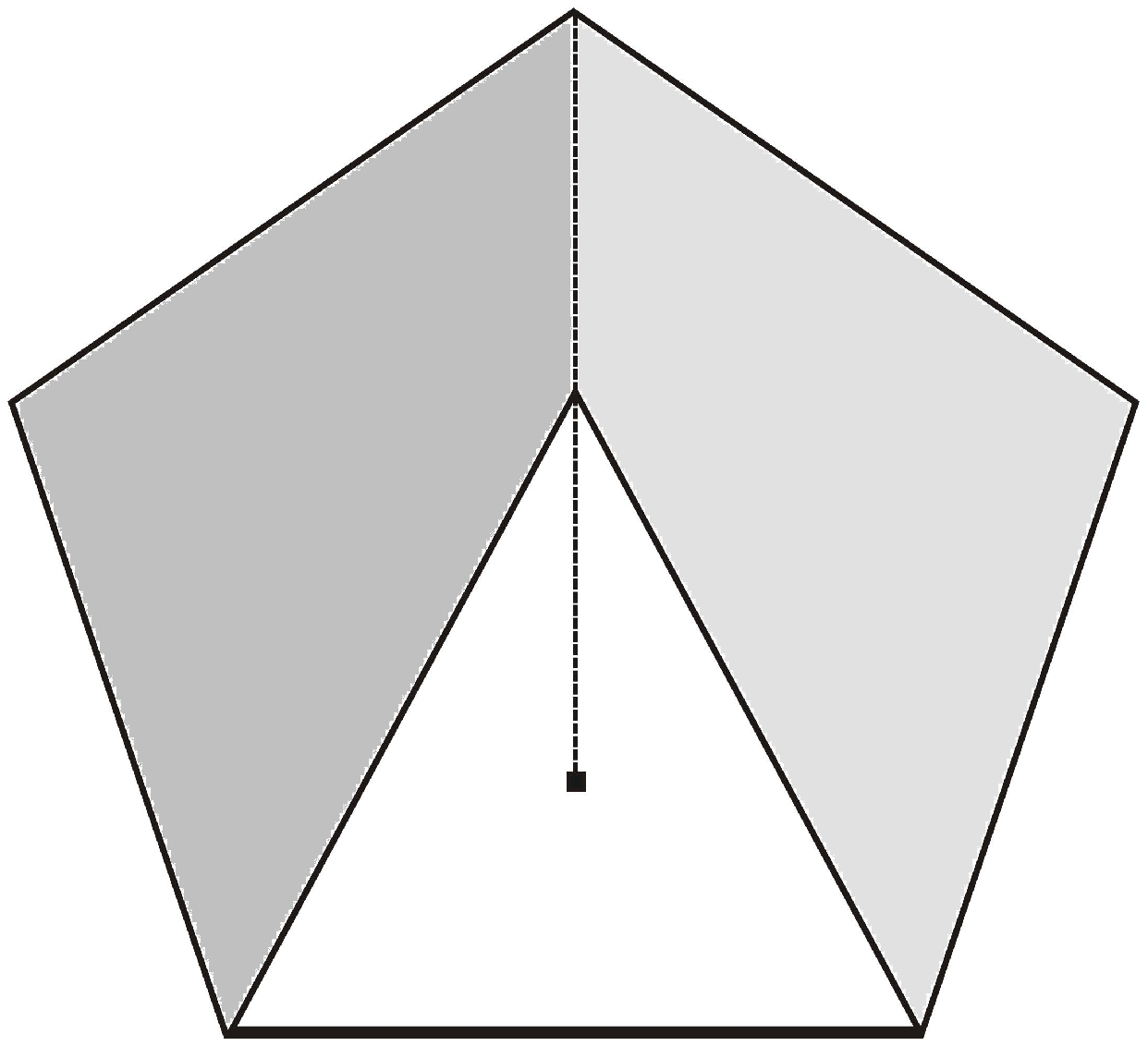}&\includegraphics[scale=0.20]{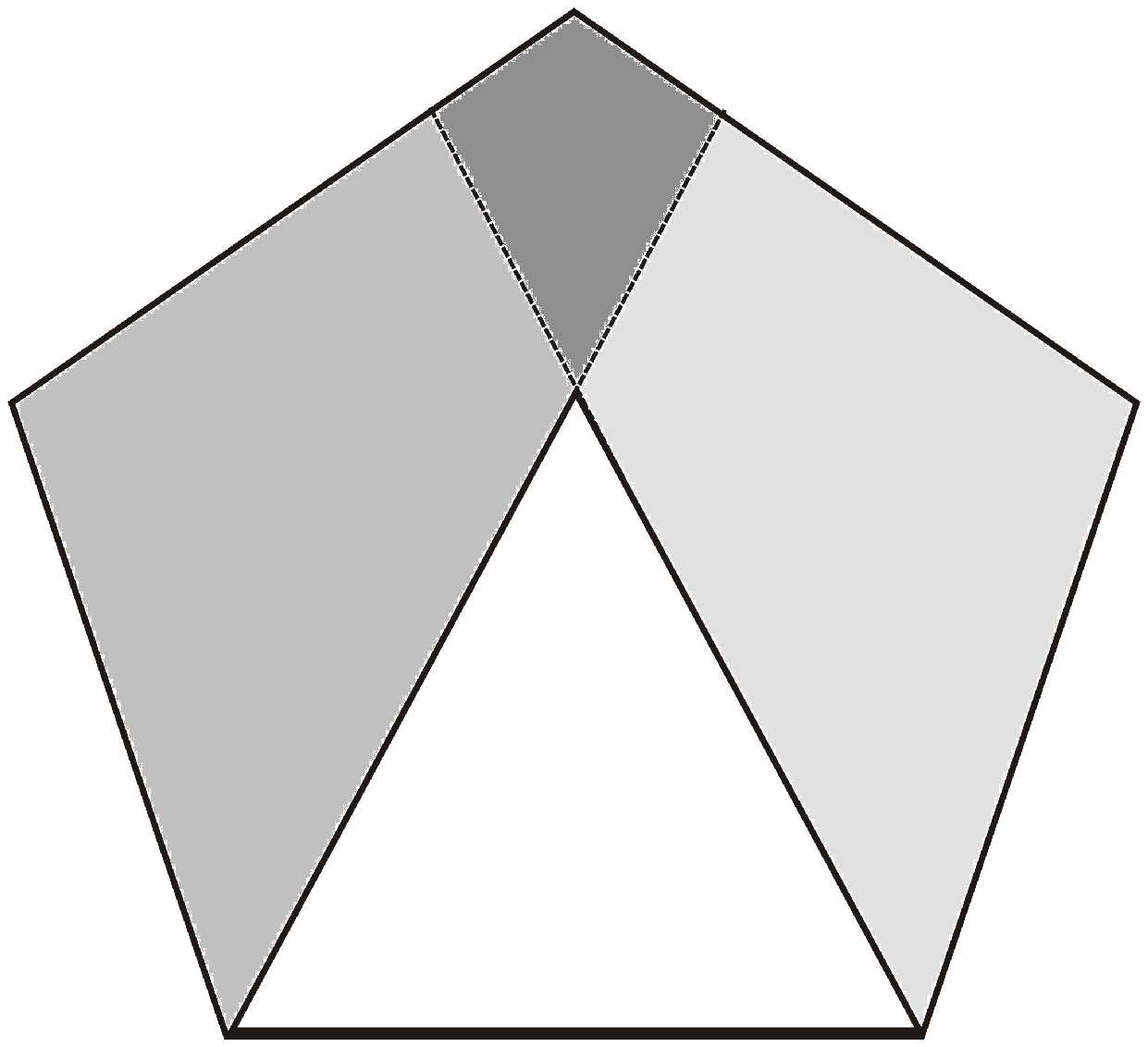}
    \end{tabular}
      \caption{{\bf{Above:}} The dot line represents the path $\rho(q)$
  followed by $\rho$ when some parameter $q$ is changed. It is
  possible to see that both $D$ and $S_k$ present singular points in its shapes. {\bf{Below-left:}}
  In a schematic picture, the pentagon represents $D$ and the triangle $S_k$, and we can draw a
  {\emph{phase diagram}} for random $k-$robustness. In addition to the separable phase, there are two
  other separated by the line which starts in the random state and passes through a vertex of the triangle.
  {\bf{Below-right:}} Analogously, the phase diagram for generalized $k-$robustness, where we can recognize
  three entangled phases, separated by lines starting at a vertex of the pentagon and passing through
  a vertex of the triangle.}\label{figGPT}
\end{figure}

Now the scenario is set and we can present the {\emph{geometric
phase transitions}}. The low dimensionality of our figures hides
some facts. The border of convex sets can be, locally, of two kinds:
curved or flat. To avoid details, let us use three-dimensional
examples: a ball and a polyhedron. In one of them, the boundary is
smooth, and there are no singular points; in the other we have edges
and vertices, which are singularities. The general picture is
neither of them. For a low dimensional example, one can think of a
cylinder, in which there are some (not too) singular points. If one
take a smooth one parameter family of states, $\rho\de{q}$, \ie a
curve on $D$, and calculate $R^k_r$ and $R^k_g$, they will usually
be smooth functions of $q$. However, singularities may appear in
these functions whenever singularities happen on the boundaries of
$S_k$ (in $R^k_r$) and of both $D$ and $S_k$ (in $R_g^k$) - see
Figure \ref{figGPT}. This is a phase transition due to the geometry
of the set of states, and it is what we call a geometric phase
transition. One should note that this abstract picture of a curve on
$D$ is just to keep things general. Important examples are given by
the time evolution of states under the effect of a Hamiltonian (for
closed systems), a master equation (open systems), or the thermal
equilibrium state as a function of temperature, or even the ground
state of a multipartite system as a function of some coupling
parameter of the Hamiltonian.

\begin{figure}[ht]\centering
  \includegraphics[width=5.2cm]{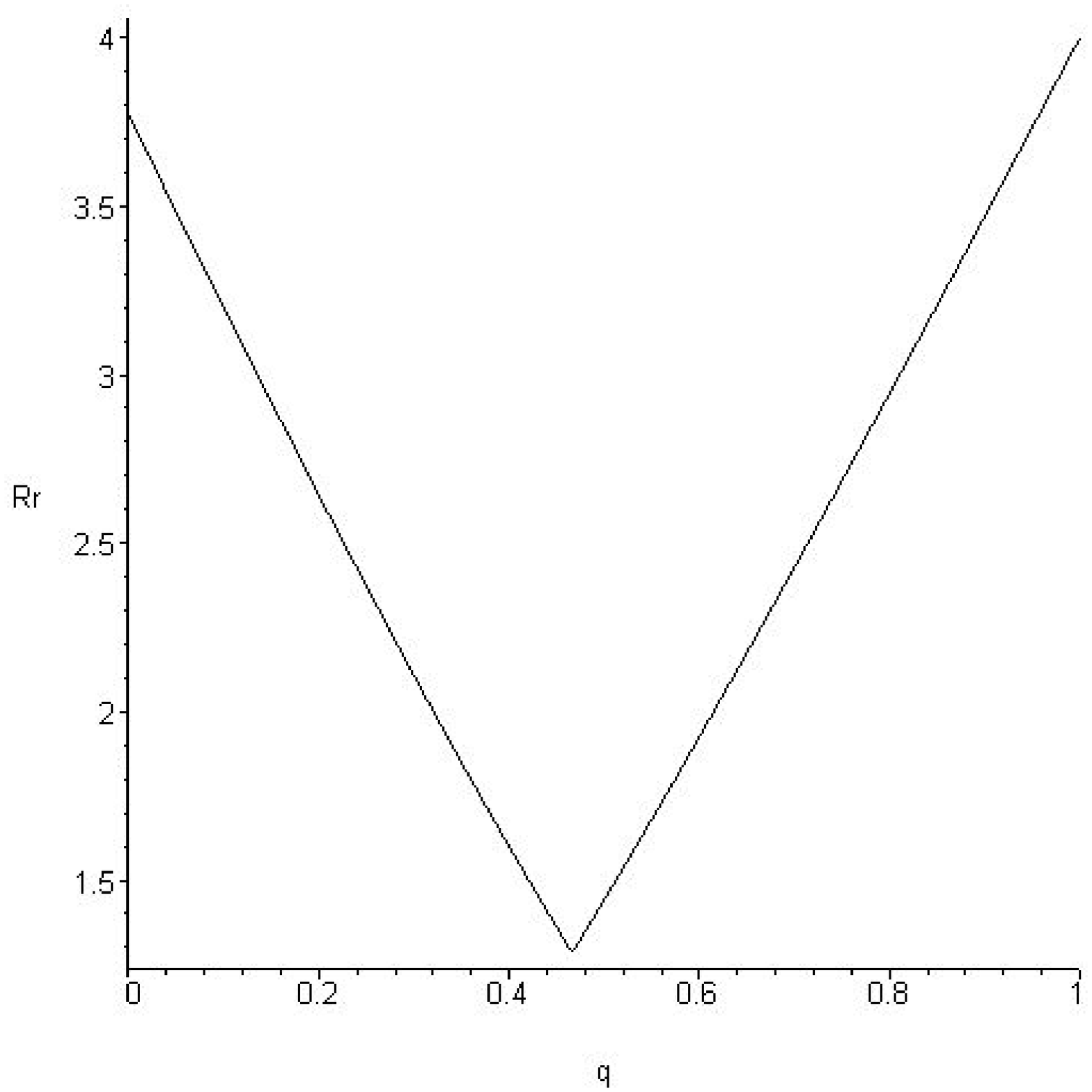}\\
  \includegraphics[width=6cm]{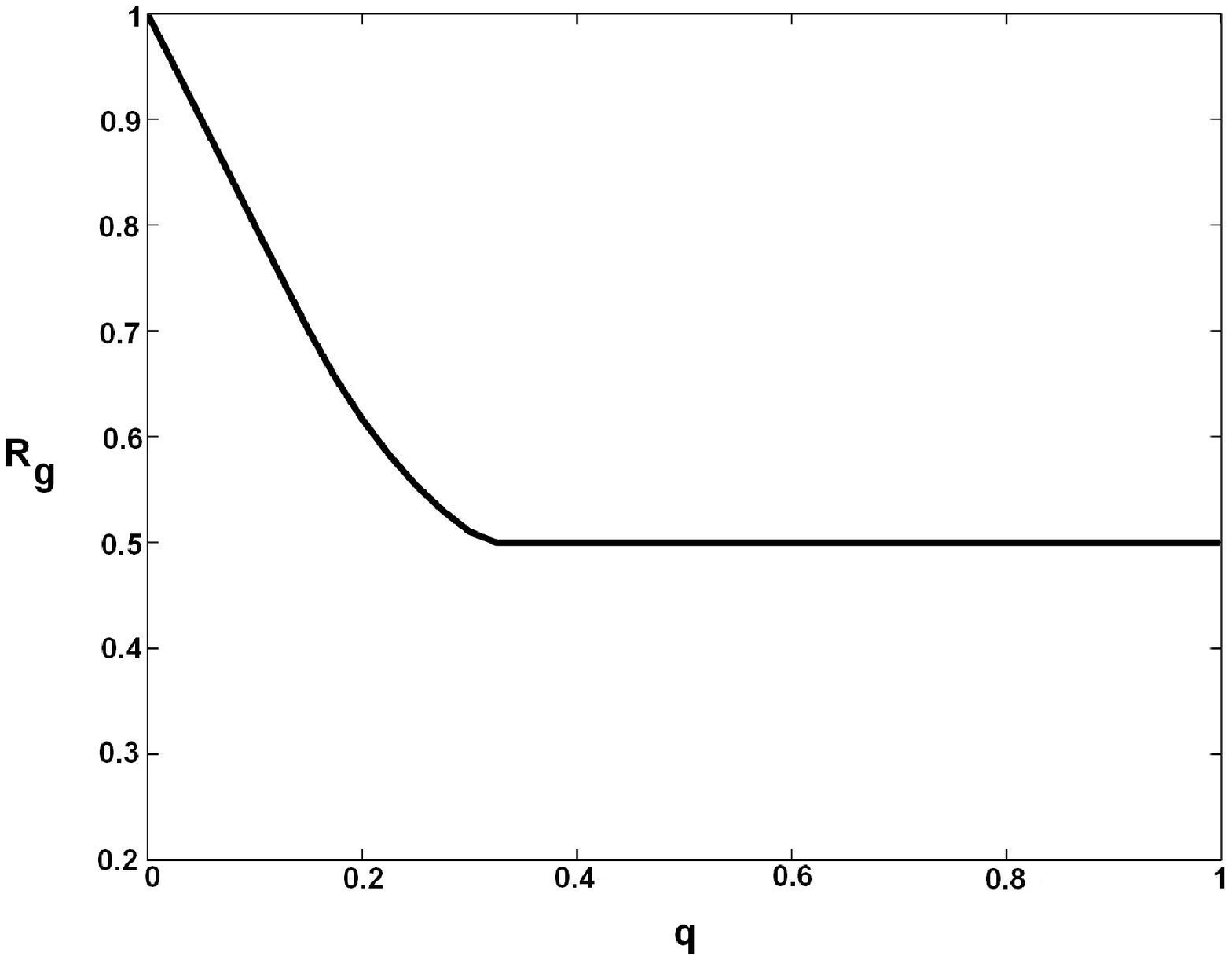}\\
  \caption{Two geometric phase transitions detected by the states
  $\rho(q)=q\ket{GHZ}\bra{GHZ}+(1-q)\ket{W}\bra{W}$,
  where $\ket{GHZ}=\frac{1}{\sqrt{2}}(\ket{000}+\ket{111})$ and
  $\ket{W}=\frac{1}{\sqrt{3}}(\ket{001}+\ket{010}+\ket{100})$. {\bf{Above:}} Random $3-$robustness,
  $R_{r}^3(\rho)$, as a function of $q$. One can see  a singularity at $q=0.47$, attesting,  thus, a singularity
  in the border of the set of 3-separable states, $S_3$ (caution: the lines in the picture are not straight lines!).
  {\bf{Below:}} Generalized $2-$robustness as a function of $q$, $R_{g}^2(\rho)$, exhibit a singularity at $q=0.33$, passing  from a decreasing function to a constant function. This constant value seems to imply parallel lines at the boundaries of $S_2$ and $D$.}\label{FigGraph}
\end{figure}

Until now we have claimed that a singularity in $R_g^{k}\de{q}$ or
in $R_r^{k}\de{q}$ is a sufficient condition to attest singularity
in $D$ or $S_k$. In fact it was possible to confirm the geometric
phase transition in some examples, two of them are displayed in
Figure \ref{FigGraph}.

As seen before the entanglement quantifiers $R_g^{k}\de{q}$ and
$R_r^{k}\de{q}$ provide information about the geometry of the
(entangled) states set. Interestingly, these functions can be
evaluated experimentally, since they are directly related to the
expected value of a physical observable. To shed light in this point
let us present the notion of entanglement witnesses \cite{HHH}: for
every $k-$entangled state $\rho$ there exists a Hermitian operator
$W^k$ (called an entanglement witness) which detects its
entanglement through:
\begin{equation}
\tr(W^{k}\rho)<0 \quad {\text{and}}
\quad \tr(W^{k}\sigma)\geq0, \quad \forall
\sigma \in S_k.
\end{equation}
A related concept is the idea of the optimal entanglement
witness\cite{Ter}, that is a witness operator $W^{k}_{opt}$ which
maximizes the value of $|\tr(W^{k}\rho)|$ when restricted by some
additional condition. It was shown \cite{Bra} that if the optimal
witness satisfy the constraint $W^{k}_{opt}<I$, then
\begin{equation}\label{RG}
R_{g}^{k}(\rho)=-\tr(W^{k}_{opt}\rho)= -\left\langle W^{k}_{opt}\right\rangle ,
\end{equation}
while if it is imposed $\tr(W^{k}_{opt})=d$, with $d$ the dimension of the total state space, then
\begin{equation}\label{RR}
R_{r}^{k}(\rho)=-\tr(W^{k}_{opt}\rho)=-\left\langle W^{k}_{opt}\right\rangle .
\end{equation}
Thus both $R_{g}^{k}$ and $R_{r}^{k}$ are given by the mean value of
some Hermitian operator $W^{k}_{opt}$ which, by the other hand, can
be linked with physical observables \cite{Ter,GHB02} (different
observables for different quantifiers, despite of our notation). The
experimental detection of $\left\langle W^{k}_{opt}\right\rangle$
was confirmed trough an optical setting to attest entanglement in
photons \cite{BEK+04}. A more detailed discussion on how to
experimentally follow such geometric phase transition is given in
the appendix. Furthermore an optimal entanglement witness
$W^{k}_{opt}$ can be viewed as a tangent hyperplane separating
$\rho$ from the set $S_k$ \cite{HHH}. It is possible to see that
both robustnesses $R_g^k$ and $R_r^k$ are not only linear functions
of $\rho$ itself, but also of $W^{k}_{opt}$. Thus a discontinuity in
$W^{k}_{opt}\de{q}$, which means a discontinuity in the family of
hyperplanes tangent to $S_k$, will cause a singularity in the
corresponding entanglement of $\rho$. As this discontinuity must be
caused by a sharp shape of $S_k$, singularities in functions
\eqref{RG} or \eqref{RR} are also necessary conditions to attest
singularities in $D$ or $S_k$.

Other entanglement quantifiers were also shown to exhibit
singularities when calculated for smooth curves $\rho(q)$ in $D$.
That is the case of the Asymptotic Relative Entropy of Entanglement
 \cite{Aud} and the Entanglement of Formation \cite{TV}. Although a
geometric interpretation of these quantifiers is not clear, these
results can be considered as predecessors of geometric phase
transitions, which suggests more research on the theme.

We finish this Letter highlighting the surprising fact that a purely
mathematical abstraction, the shape of the set of quantum states, can be
directly tested by real experiments. Despite it being a
philosophical question that deserves further
investigation, a better comprehension of the geometry behind
entanglement can help on understanding several physical phenomena.
In fact, the understanding of quantum correlations is one of the
greatest challenges to the contemporary physics, with a wide range of applications.

\begin{acknowledgments}
It is a pleasure to thank R. Dickman, F. Brochero, M. B. Plenio, and
V. Vedral for enlightening discussions, and F. Tenuta for helping us
with the figures. The authors also thank T.G. Mattos and R. Falc\~ao
for useful comments on a previous version of this manuscript.
Financial support from Brazilian agency CNPq is also acknowledged.
\end{acknowledgments}

{\emph{Appendix-}} It was argued in the main text that optimal
entanglement witnesses (OEW's) can be experimentally implemented
and, thus, used to study the geometry behind entanglement. However,
in order to do that, one must firstly have a feasible way of
determining an OEW to general states and, in a second moment, one
must know how to implement this OEW in a real experimental setting.
In this appendix we aim to show some methods to reach this end.

First of all let us discuss how to find an OEW with the prior
knowledge of the state $\rho$. If $\rho$ is a pure state, \ie
$\rho=\ket{\psi}\bra{\psi}$, then an analytical method can be used.
An OEW, $W_{opt}^{k}$, of $\rho$ is given by\cite{BEK+04,Ter,Wei}
\begin{equation}
W_{opt}^{k}=\lambda I-\ket{\psi}\bra{\psi},
\end{equation}
with $$\lambda=\max_{\ket{\sigma} \in S_k}
|\braket{\psi}{\sigma}|^{2},$$
 and $I$ being the identity matrix. A way to compute $\lambda$ is already known\cite{BEK+04}.
 For mixed states (with the exceptions of two qubits and qubit-qutrit cases, where the Peres partial
 transposition criterion is decisive and can be interpreted in the entanglement witness context),
 there is no analytical way of finding it. However,  there exist efficient numerical algorithms to
 approximate it\cite{Bra04a,Bra04b,Eisert}.

In this sense, the way we envisage to follow experimentally such
geometric phase transition begins with a procedure to prepare states
depending on one parameter $q$, $\rho\de{q}$. For each fixed $q$
value, the experimentalist proceeds a tomographic experiment. Then,
one of the numerical algorithms should be executed to find an OEW
for this state, $W_{opt}\de{q}$. The next step involves to measure
$\mean{W_{opt}\de{q}}$. This must be repeated for other values of
$q$ and, in this way, graphics like the ones showed in the text can
be generated experimentally.

This can be criticized as being a very indirect way of measuring
something. But one must remember that any physical experiment is
guided by a theory that one wants to put in check. On the other
hand, this is just a general procedure that can be much easier in
specific cases. In many situations, the same EW is optimal for some
range of value of $q$, and if one experimentally believes in the
state that is being generated, the comparison of two specific
entanglement witnesses can be enough to show a geometric phase
transition.

\end{document}